\documentclass[aip, reprint]{revtex4-2}

\usepackage{amsmath}
\usepackage{amssymb}
\usepackage{graphicx}
\usepackage{dcolumn}
\usepackage{bm}
\usepackage[mathlines]{lineno}

\usepackage[utf8]{inputenc}
\usepackage[T1]{fontenc}
\usepackage{mathptmx}
\usepackage{etoolbox}

\usepackage{enumitem}
\usepackage{chemformula}

\usepackage{microtype}

\usepackage[colorlinks, allcolors=blue]{hyperref}
\hypersetup{pdftitle={aenet-PyTorch}, pdfauthor={Jon López-Zorrilla}}

\draft 

\begin{document}

\title{ænet-PyTorch: a GPU-supported implementation for machine learning atomic potentials training}

\author{Jon López-Zorrilla}
\email{jon.lopezz@ehu.eus}
\affiliation{Physics Department, University of the Basque Country (UPV/EHU), Leioa, Basque Country, Spain}

\author{Xabier M. Aretxabaleta}
\affiliation{Physics Department, University of the Basque Country (UPV/EHU), Leioa, Basque Country, Spain}

\author{Inwon Yue}
\affiliation{Department of Chemical Engineering, Columbia University, New York, NY 10027, United States}

\author{Iñigo Etxebarria}%
\affiliation{Physics Department, University of the Basque Country (UPV/EHU), Leioa, Basque Country, Spain}
\affiliation{EHU Quantum Center, University of the Basque Country (UPV/EHU), Basque Country, Spain}

\author{Hegoi Manzano}%
\affiliation{Physics Department, University of the Basque Country (UPV/EHU), Leioa, Basque Country, Spain}

\author{Nongnuch Artrith}
\email{n.artrith@uu.nl}
\affiliation{Materials Chemistry and Catalysis, Debye Institute for Nanomaterials Science, Utrecht University, Utrecht, The Netherlands}

\date{\today}

\begin{abstract}
In this work, we present ænet-PyTorch, a PyTorch-based implementation for training artificial neural network-based machine learning interatomic potentials. Developed as an extension of the atomic energy network (ænet), ænet-PyTorch provides access to all the tools included in ænet for the application and usage of the potentials. The package has been designed as an alternative to the internal training capabilities of ænet, leveraging the power of graphic processing units to facilitate direct training on forces in addition to energies. This leads to a substantial reduction of the training time by one to two orders of magnitude compared to the CPU implementation, enabling direct training on forces for systems beyond small molecules. Here we demonstrate the main features of ænet-PyTorch and show its performance on open databases. Our results show that training on all the force information within a data set is not necessary, and including between $10\%$ to $20\%$ of the force information is sufficient to achieve optimally accurate interatomic potentials with the least computational resources.
\end{abstract}

\maketitle

\section{Introduction}

In recent years, machine learning methods have gained popularity as methods to simulate large complex systems, due to their ability to predict properties of materials with high accuracy and low computational cost. In particular, machine learning interatomic potentials\cite{mlp:behler2007generalized} (MLPs)  are data-driven methods that allow the prediction of energies and forces of atomic structures with precision similar to that of the scheme used to generate the reference data but several orders of magnitude faster. Reference data is usually obtained by first-principles calculations, such as density functional theory\cite{dft:burke2012perspective,dft:fiedler2022deep} (DFT) for bulk systems or post-Hartree-Fock methods for molecular ones\cite{cc:bishop1987coupled,cc:smith2019approaching}. Once trained, these potentials can be employed in conjunction with other advanced simulation techniques like molecular dynamics\cite{aenet:chen2021aenet} (MD), Monte Carlo\cite{mc:chen2021microstructure,mc:nagai2020self,mc:tirelli2022high} (MC), evolutionary algorithms\cite{aenet:artrith2018constructing}, etc. MLPs have been used with great success in a wide variety of fields including physics, chemistry and industry, in applications such as computing phonon properties\cite{phonon:bartok2018machine,phonon:rowe2018development}, studying phase diagrams\cite{phase:chew2022phase,phase:kruglov2019phase}, or predicting properties and structures of crystals and molecules\cite{discover:gubaev2018machine,discover:pilania2013accelerating,discover:podryabinkin2019accelerating,discover:schmidt2017predicting}.

Multiple MLP approaches have been proposed in the literature, some examples include artificial neural network-based potentials (ANN-based MLPs) \cite{mlp:behler2011neural,mlp:behler2017first,mlp:behler2014representing}, gaussian approximation potentials \cite{mlp:bartok2010gaussian,mlp:szlachta2014accuracy,mlp:bartok2017machine} kernel-based methods \cite{krr:john2017many, agni:botu2017machine, agni:botu2015learning}, message-passing networks \cite{mess:gilmer2017neural, mess:xue2021reaxff, mess:schutt2017quantum}, or spectral neighbor analysis potentials \cite{snap:thompson2015spectral,snap:wood2018extending} among many others. In this case, our focus lies on the first group, ANN-based MLPs, which is based on one of the oldest and most studied methods in machine learning. The main drawback of MLPs is that the calculations to compile the reference data set and training the potential is time- and resource-intensive. Several fits are usually needed, i.e., the training needs to be repeated several times, first to select the optimal hyperparameters, and then to refine the MLP with techniques such as active or on-the-fly learning \cite{fly:botu2015adaptive,fly:jacobsen2018fly,fly:wang2020lithium,fly:jinnouchi2019fly}. The simplest of the approaches relies on training only on the energies of the reference structures. However, this approach usually leads to poor predictions of forces, which are the negative gradient of the predicted potential energy with respect to the atomic coordinates, and are of foremost importance to achieve long stable MD simulations\cite{mlp:unke2021machine}. Therefore efficient techniques are required to include force information in the training  in addition to the reference potential energies, which is far more computationally demanding.

As in most fields concerning machine learning, Graphics Processing Units (GPUs) provide the best solution to this problem. Some work has already been done in recent years in this direction, leading to updates of codes for training MLPs to include GPU support. For instance that is the case of ANI \cite{soft:gao2020torchani}, AMP\cite{soft:khorshidi2016amp} or deepMD \cite{soft:wang2018deepmd}, to name but a few. Here we present an extensive update of the atomic energy network (ænet) \cite{aenet:artrith2016implementation} code to allow MLP training on GPUs. ænet has proved to be efficient, especially when handling systems with many species. Our approach is simple: by using a well-known ML framework, PyTorch \cite{soft:paszke2019pytorch}, we replace the training process of ænet while keeping full compatibility with all the other ænet resources. In this paper, we describe the main characteristics of our code called ænet-PyTorch and show its potential to train on both system energy and atomic forces. We also show that training on all the forces of a database is redundant by testing the code on several open databases. The code will be available on GitHub\cite{pytorch_aenet} as free, open-source software.

\subsection{Machine learning potentials}

The main reason for the success of ANN-based MLPs is that once trained they can be used to predict energies and forces of new structures independently of the number of atoms, yet they are limited to the chemical species present in the reference data. This is achieved by partitioning the energy of the system into atomic contributions:
\begin{equation} \label{eq0:E_ann}
    E^\text{ANN} (\{\sigma^{(i)}\}) = \sum_{i=1}^{N_\text{atom}} E_i(\sigma^{(i)})
\end{equation}
A neural network is trained for each chemical element present in the reference data, and the contribution of each atom $(E_i)$ is then evaluated using the network specific to its element. It is assumed that the contribution of each atom $i$ depends only on its local environment (denoted as $\sigma^{(i)}$ in the previous equation). By numerically describing those environments, via the so-called atomic fingerprints or descriptors and training on the total energy of the system, the resulting potential can be generalized independently of the number of atoms. Several ways of representing atomic environments can be found in the literature\cite{review:unke2021machine,descr:musil2021physics,review:yaghoobi2022machine,soap:bartok2013representing,soap:de2016comparing}, all of them satisfying a set of conditions regarding symmetry with respect to the exchange of equivalent atoms, rotations and translations of the structures, and smoothness of the descriptor functions.

The training is an optimization process, so the first step is to define the objective function to be minimized. There are two main approaches: training only on energies, or including also information about the forces acting on each atom. In any case, the objective function (also known as the loss function) is usually the root mean squared error (RMSE) of the ANN output compared to the reference value. We define the loss function ($\mathcal{L}_{EF}$) as a weighted sum of both energy and force errors
\begin{equation} \label{eq1:loss_ef}
    \mathcal{L}_{EF} = (1-\alpha) \mathcal{L}_E + \alpha\mathcal{L}_F
\end{equation}
where $\alpha$ is a free weight parameter. RMSEs for energy ($\mathcal{L}_{E}$) and forces ($\mathcal{L}_{F}$) are defined as
\begin{equation}
    \mathcal{L}_E = \sqrt{\frac{1}{N_\text{struc}} \sum_{i=1}^{N_\text{struc}} \left( E_i^\text{ANN} - E_i^\text{REF} \right)^2}
\end{equation}
and
\begin{equation}
    \mathcal{L}_F = \sqrt{\frac{1}{\sum_{i=1}^{N_\text{struc}} 3{N_{\text{atom},i}}} \sum_{i=1}^{N_\text{struc}}  \sum_{j=1}^{N_{\text{atom},i}} \left( \mathbf{F}_{i,j}^\text{ANN} - \mathbf{F}_{i,j}^\text{REF} \right)^2}
\end{equation}
where $N_\text{struc}$ is the number of structures in the database and $N_{\text{atom},i}$ the amount of atoms in the $i$th structure. Here the superscripts ANN and REF denote the output of the MLP and the reference value, respectively. For the sake of simplicity, the force error is normalized per atom in the whole set, instead of having the average of force error per structure. That is to say, we consider each force vector as an independent training example\cite{soft:singraber2019parallel}. The outputs of the MLP ANNs are the energy contributions of each atom in Eq. \ref{eq0:E_ann}, and the forces acting on each atom can be computed by taking the gradient of the total energy with respect to the coordinates of the atoms ($\{\mathbf{R}_k\}$):
\begin{equation}
    \mathbf{F}^\text{ANN}_k = -\nabla_k E^\text{ANN}(\{\sigma^{(i)}\})
    = -\sum_{i=1}^{N_\text{atom}} \sum_{n=1}^{N_\text{descr}} \frac{\partial E_i}{\partial \sigma_n^{(i)}} \frac{\partial \sigma_n^{(i)}}{\partial \mathbf{R}_k}
\end{equation}

Training the network means finding an optimal set of weights and biases that minimize the loss function. It is a common practice to include one more term in the loss function which is called regularization or weight decay. It is intended to avoid overfitting, i.e., finding a solution that minimizes the error for the training examples but does not generalize to examples outside the reference data set\cite{mlp:miksch2021strategies}. Here we consider the L2 regularization\cite{soft:krogh1991simple}, introduced in the training as an extra term in the loss function
\begin{equation} \label{eq2:loss_reg}
    \mathcal{L} = \mathcal{L}_{EF} + \frac{1}{2}\lambda \sum_i w_i^2
\end{equation}
where $\{w_i\}$ is the set of all parameters to be fitted during training, and is $\lambda$ a weighting parameter. In the following, that $\lambda$ parameter will be referred to as the regularization or weight decay term.

In practice, in each training step the loss function in Eq. \ref{eq2:loss_reg} is not evaluated for the whole set of training examples, but rather for a subgroup of them at a time. This group, or batch, is used to approximate the gradient of the loss function, and to update the parameters in the network in each step.

\section{ænet-PyTorch}

\subsection{ænet}

ænet\cite{aenet:artrith2016implementation} provides a tool for generating, testing, and applying machine learning interatomic potentials based on the Behler-Parrinello method \cite{mlp:behler2007generalized}, entirely written in modern Fortran 2003.  Currently, it includes two different descriptors: the original Behler atom-centered symmetry functions  \cite{mlp:behler2011atom,mlp:behler2007generalized} and the Chebyshev descriptors \cite{mlp:artrith2017efficient} by Artrith et al. Nonetheless, it is worth noting that this update of the code is independent of the descriptor, so any future addition of new descriptors in the original ænet code would be compatible with ænet-PyTorch.

ænet also includes the utilities to employ the MLPs in real applications, for example, in molecular dynamics simulations\cite{aenet:chen2021aenet} (using the interface with LAMMPS\cite{LAMMPS} or TINKER\cite{tinker}), in path integral molecular dynamics\cite{pimd:kimizuka2022artificial}, or in any of the capabilities provided in the Atomic Simulation Environment \cite{soft:larsen2017atomic} Python package. Our new program has been designed as an extension to the original ænet code, and all of these utilities that ænet provides can be readily used with the potentials generated by ænet-PyTorch.

Including force information in the training increases the transferability of the potentials, improving the force prediction, and finally enhancing the stability of MD simulations. Nevertheless, despite ænet's parallel implementation's efficient scaling on CPUs, training on both energies and forces has been limited to simple systems with a small number of atoms due to its significant requirements for computational time and memory. That is the main reason why we consider this GPU implementation through PyTorch necessary so that the force training becomes feasible for complex systems within our ænet framework.

\subsection{ænet-PyTorch implementation}

\begin{figure*}[htb]
    \centering
    \includegraphics[width=0.8\textwidth]{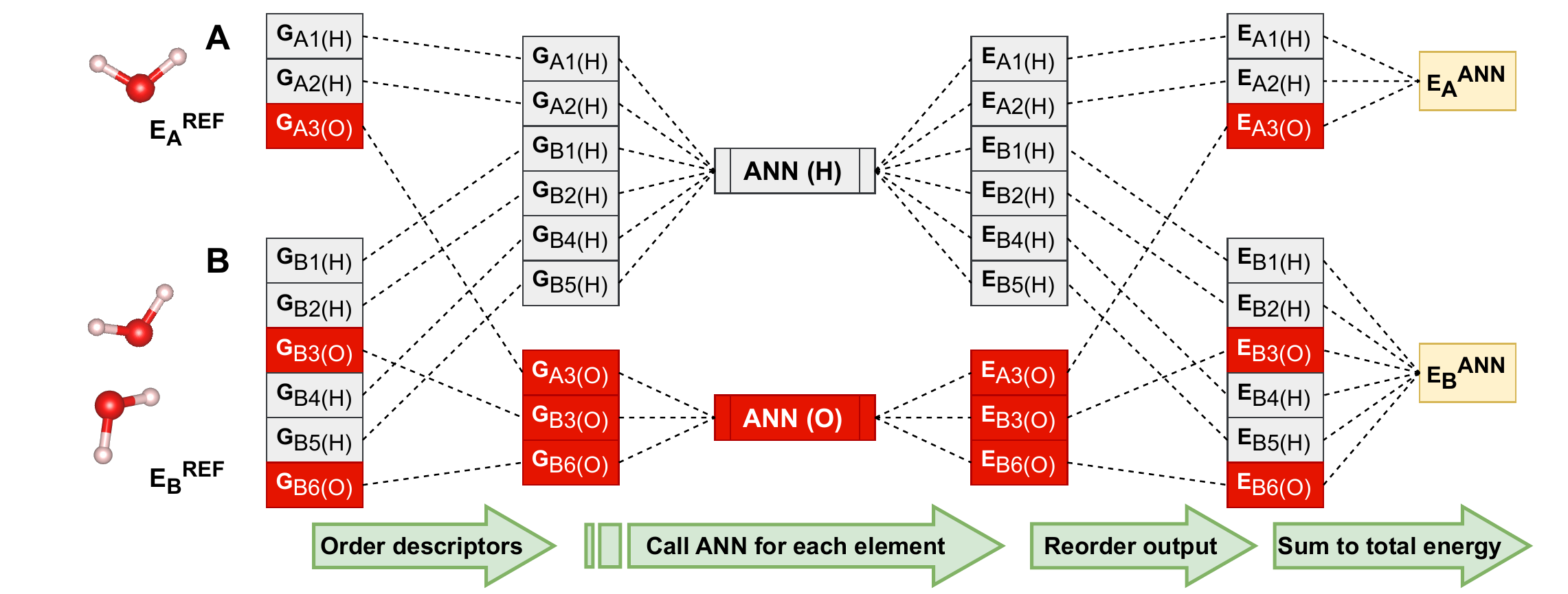}
    \caption{In this example of the workflow of the code, the data set is composed of two structures: A with one water molecule and B with a cluster of two water molecules. First, all the descriptors of the hydrogen atoms of both A and B structures are grouped in the same tensor, and so are the ones for the oxygen atoms. A single call to the neural network for the hydrogen (oxygen) atoms computes the atomic contribution of all of the atoms of that element. Then, the ordering process is reverted and the atomic energies are grouped per structure again. Finally, the contributions of each atom are summed to obtain the total energy of each structure, $E^\text{ANN}_A$ and $E^\text{ANN}_B$. All these operations are implemented via built-in PyTorch routines. }
    \label{fig0:implementation}
\end{figure*}

The principal novelties included in the training scheme of ænet-PyTorch are the capability to train MLPs using GPU in addition to CPU cores, and training both on reference energies and atomic forces. ænet-PyTorch also provides users with easy access to various optimization algorithms and overfitting prevention techniques such as dropout and batch-normalization layers, all available within the PyTorch framework.All of these are easily implementable and customizable for users.

PyTorch is a free and open-source machine learning framework based on the TorchLib library with a Python interface. It contains routines to enable efficient training of deep neural networks with CPU and GPU support, via the C++ and CUDA-based code, respectively. In contrast to other PyTorch-based implementations of MLPs, which usually rely on custom CUDA or C++ modules, this one is written completely in Python using only PyTorch's built-in functions. More specifically, GPUs are best suited for tensor operations, so most operations must be expressed as such to achieve optimal performance.

The optimization strategy (depicted in Fig. \ref{fig0:implementation}) involves grouping all atoms of the same species together before training. This allows a single network call to predict the energies of all atoms of that type. PyTorch handles the parallel computation of atomic energies and forces internally. The process of grouping all the atoms and later reordering the resulting atomic energies and forces is also performed using only PyTorch's built-in routines, so it is also a parallel process. As we will see in the next section, this idea leads to great scaling of the code, particularly on GPU.

For a detailed explanation of the capabilities and options of the code, the user is referred to the documentation of the code which can be found online in the GitHub repository (\url{https://github.com/atomisticnet/aenet-pytorch}) together with the code and an example of its usage. However, some of them are worth noting, and they will be tested in the following. One of the main concerns when developing MLPs is the resources needed to train forces, both in terms of computational time and memory. It must be noted that with our implementation we decided to sacrifice memory resources to speed up the training several orders of magnitude. However, some options have been included to reduce those memory requirements for large data sets, both CPU and GPU memory (called RAM and VRAM, respectively):

\begin{itemize}[align=left]
    \item \textbf{GPU (gpu)}: This mode stores all the data needed on the GPU memory. Despite being the fastest way, storing all the data in the GPU device might be problematic for large data sets.
    \item \textbf{GPU (cpu)}: If enough RAM is available but that of the GPU is limited, the data can be all stored in the CPU RAM. Then for each training iteration only the required data is moved to the GPU.
    \item \textbf{CPU}: All calculations are done using CPU cores, without GPU support.
\end{itemize}

All three approaches, GPU (gpu), GPU (cpu) and CPU, will be tested in the following sections.

\section{ænet-PyTorch scaling}

In order to test the performance of ænet-PyTorch, we will replicate already published calculations using this new code and compare the results with those published by their original authors. Let us first consider the database used in the first ænet code release \cite{aenet:artrith2016implementation}, formed by 7815 structures belonging to different phases of bulk titanium dioxides (\ch{TiO_2}). In this section, we will use this set of calculations to check the performance of the code for energy and force training. All MLP training would usually require the selection of an architecture for the ANN, but in our case, we will be using the one that the authors proposed and proved to be optimal: $48-15-15-1$ architecture using the Chebyshev polynomials as descriptors.

\subsection{Energy training}

\begin{table}
\caption{ List of parameters yielding the best training results for the \ch{TiO_2} database using only energy information: batch size (BS), learning rate (LR), weight decay (WD), and root mean squared error of the energy of the validation set (RMSE). The RMSE is given in meV/atom.}
\begin{ruledtabular}
\begin{tabular}{rrrrr}
\mbox{Method} & \mbox{BS} & \mbox{LR} & \mbox{WD} & \mbox{RMSE}\\
\hline
\mbox{Adagrad} & 64 & $10^{-1}$ & $10^{-4}$  &  4.221 \\
\mbox{Adamw} & 128 & $10^{-4}$ & $10^{-5}$  &  4.434 \\
\mbox{Adamw} & 256 & $10^{-4}$ & $10^{-5}$  &  4.387 \\
\mbox{Adamax} & 512 & $10^{-3}$ & $10^{-5}$  &  4.310 \\
\mbox{Adamax} & 1024 & $10^{-3}$ & $10^{-4}$  &  4.877 \\
\end{tabular}
\end{ruledtabular}
\label{tab1:parameters}
\end{table}

As we have already stated, this implementation allows the usage of all the tools in PyTorch, starting with all the optimization algorithms. Therefore, we first aim to select the most appropriate set of hyperparameters for the training. This includes the optimization method, batch size, learning rate, and weight decay. To do so, we consider all the variants of the Adam optimizer \cite{adam:kingma2014adam,adam:zeiler2012adadelta,adam:loshchilov2017decoupled} available in PyTorch, batch sizes ranging from $64$ to $1024$, learning rates from $10^{-6}$ to $10^{-1}$, and weight decays from $10^{-5}$ to $10^{-2}$. This results in a total of 600 training processes.

The models are evaluated based on their accuracy on the independent validation set, and the best results are displayed in table \ref{tab1:parameters}. All calculations are performed for $10^4$ training iterations to ensure that those with lower learning rates have also converged.

\begin{figure}[htb]
    \centering
    \includegraphics[width=0.49\textwidth]{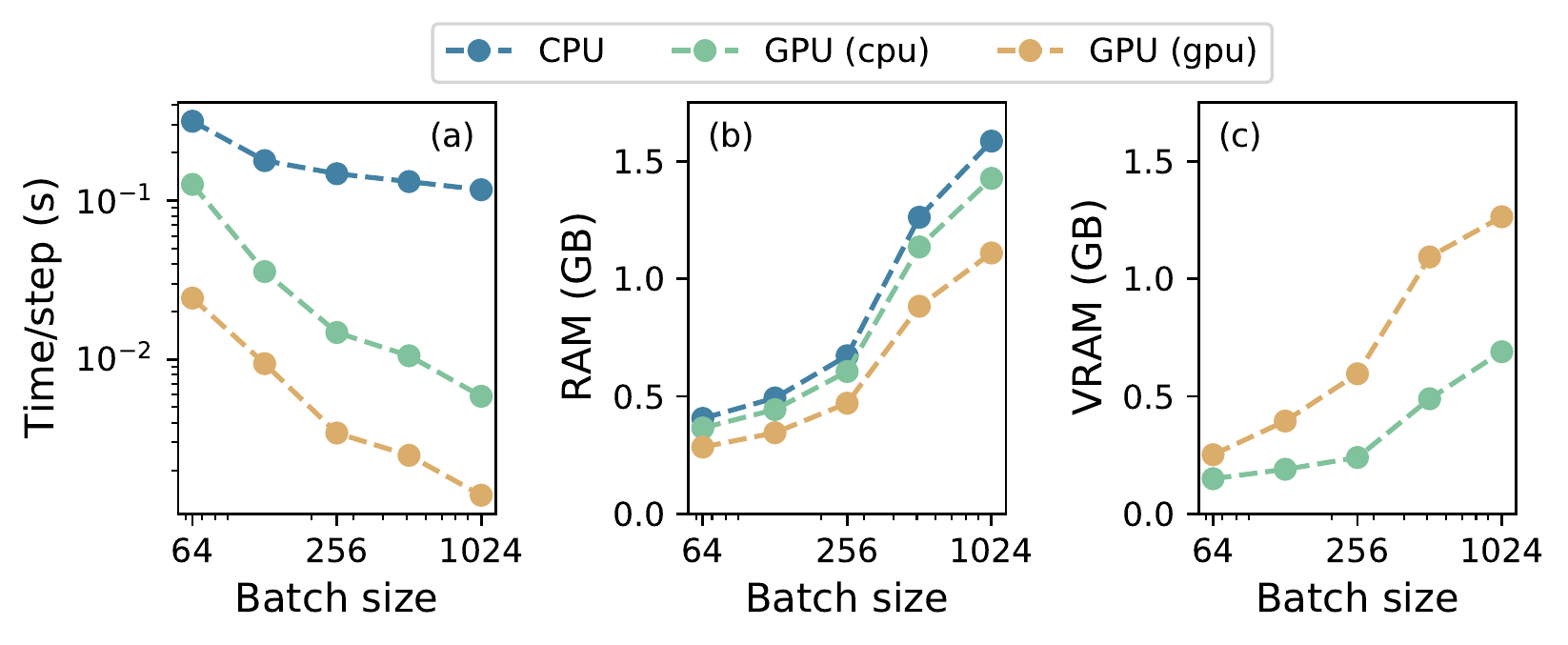}
    \caption{Resources used for training the neural network on the energies of the \ch{TiO_2} database as a function of the batch size: (a) Time per training step, (b) CPU RAM memory and (c) GPU VRAM memory.}
    \label{fig1:E_scaling}
\end{figure}

It has already been discussed that the feature that makes ænet-PyTorch efficient is the grouping of atoms of the same species before training. This has some implications: the time needed per training step considerably decreases with the increase of the batch size, but it comes at the cost of more memory.  Fig. \ref{fig1:E_scaling} shows the time needed per training step for different batch sizes. The number of steps needed to achieve convergence depends on each specific case, notwithstanding, the time per step is a good indicator of the scaling of the code. For each value, the best combination of method, learning rate, and regularization have been selected based on the results shown in table \ref{tab1:parameters}. The time needed for training considerably decreases when using GPU. The improvement ranges from one to two orders of magnitude, increasing with the batch size. On the other hand, the memory used also increases with the batch size, so in some cases, it would be better to save the data set information in the CPU RAM, so as to find a trade-off between time- and memory- efficiency. In that case, the speedup is still considerable, while keeping the GPU memory (which is usually the limiting computational feature) more moderate. However, these problems rarely arise from calculations involving only energies.

These results show that energy training with ænet-PyTorch is efficient even on CPUs, considering that the CPU computations here are performed using only 2 CPU cores. As a reference, this training time is similar to that needed in the original ænet to obtain a similar error training with 16 processors. This comparison is not completely fair, since ænet uses an optimization algorithm much slower, the limited memory BGFS algorithm.

\subsection{Force training}

\begin{figure}
    \centering
    \includegraphics[width=0.49\textwidth]{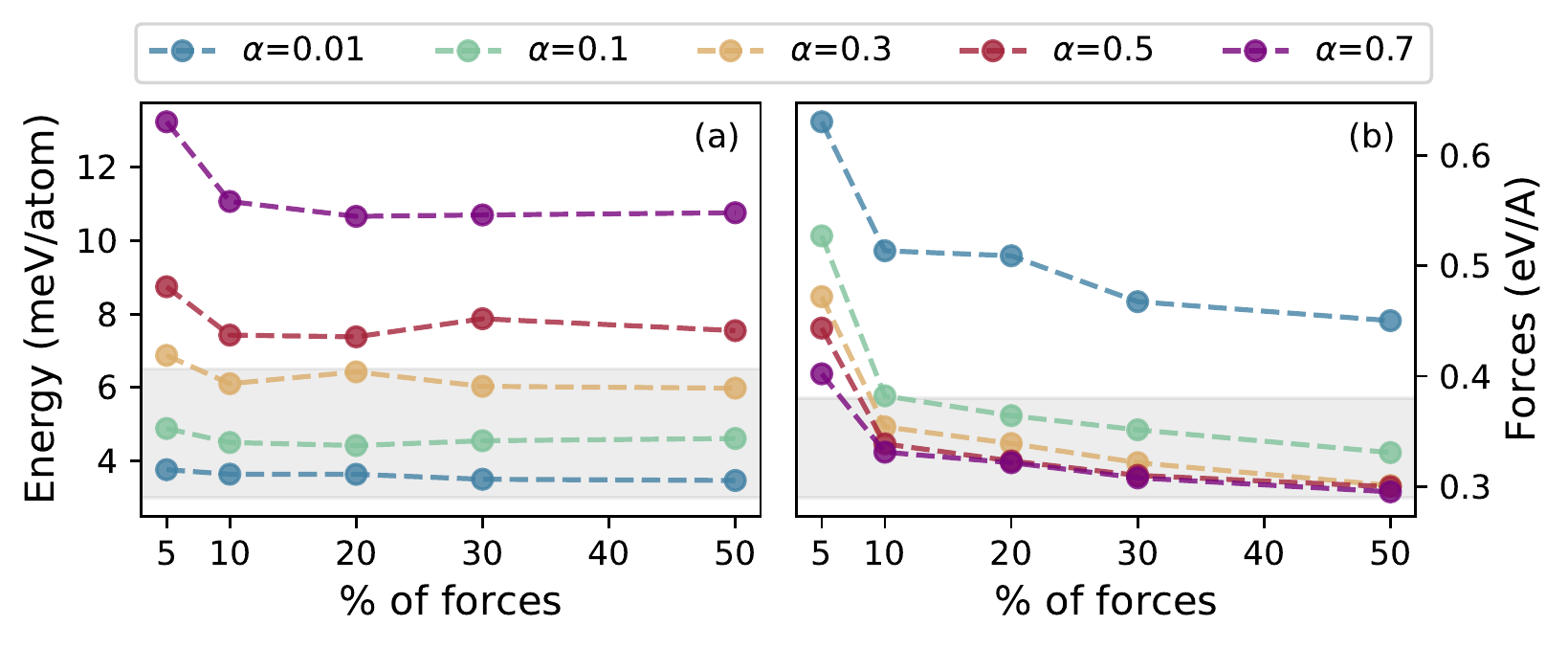}
    \caption{ Validation error of (a) energy and (b) forces as a function of the percentage of force information included in the training for different values of the weight parameter $\alpha$ for the \ch{TiO_2} database. The grey region is meant to guide the eye towards the set of parameters that we consider optimal.}
    \label{fig2:F_alpha_percent}
\end{figure}

Let us now consider the new feature introduced with this code, i.e., training on atomic forces in addition to energies. In this case, we will use the optimal training hyperparameters selected for energy training, and we will focus on two other parameters influencing force training: the $\alpha$ parameter from Eq. \ref{eq1:loss_ef} weighting the energy and force RMSEs in the loss function and the fraction of structures with force information. Regarding the former, the $\alpha$ parameter is bounded from 0 to 1, i.e., from pure energy training to only training on forces. As for the latter, some works in the field already suggest that including the forces of all the atoms of every structure in the reference set is not necessary to reach accurate predictions: Singraber et al. used from $0.41\%$ to $4.1\%$ \cite{soft:singraber2019parallel}, and Artrith et al. stated that $10\%$ was enough \cite{for:artrith2012high,al:artrith2013neural}. The results of our tests are depicted in Fig. \ref{fig2:F_alpha_percent}, where both the energy and force RMSEs are displayed as a function of the fraction of structures whose forces have been used in the training. These structures are randomly selected from the whole training set.

First, the $\alpha$ parameter determines the balance between the accuracy of energies and forces. Lower values result in lower error of energies but higher error of forces, while higher values result in the opposite. For MD simulations, getting accurate forces is more important to ensure the stability of the simulations, whereas for Monte Carlo calculations the energy is the main issue. Therefore, we need to find a balance between accuracy in energies and forces depending on the task at hand. It is important to note that the introduction of forces helps to generalize the prediction of energies in regions that are not included in the training examples\cite{soft:gao2020torchani,review:unke2021machine}, so in most cases the introduction of forces will be beneficial to some extent.

In our current example, the prediction error for forces decreases with increasing alpha, but for values above $0.1$ the performance improves very little. In the case of the energies, all the models with $\alpha$ below $0.3$ predict energies with an accuracy on the order of $~1$ meV/atom. Thus, we can conclude that $\alpha \in (0.1,0.3)$ is the optimal range of values for a general-purpose potential for this system.

As shown in Fig. \ref{fig2:F_alpha_percent}(b) the error of the forces does indeed decrease when adding more force information to the training set. However, this improvement diminishes when the percentage of forces included is high enough. This means that including from $10\%$ to $20\%$ is sufficient to achieve results close to the best performance that the model can provide.

\begin{figure}[htb]
    \centering
    \includegraphics[width=0.49\textwidth]{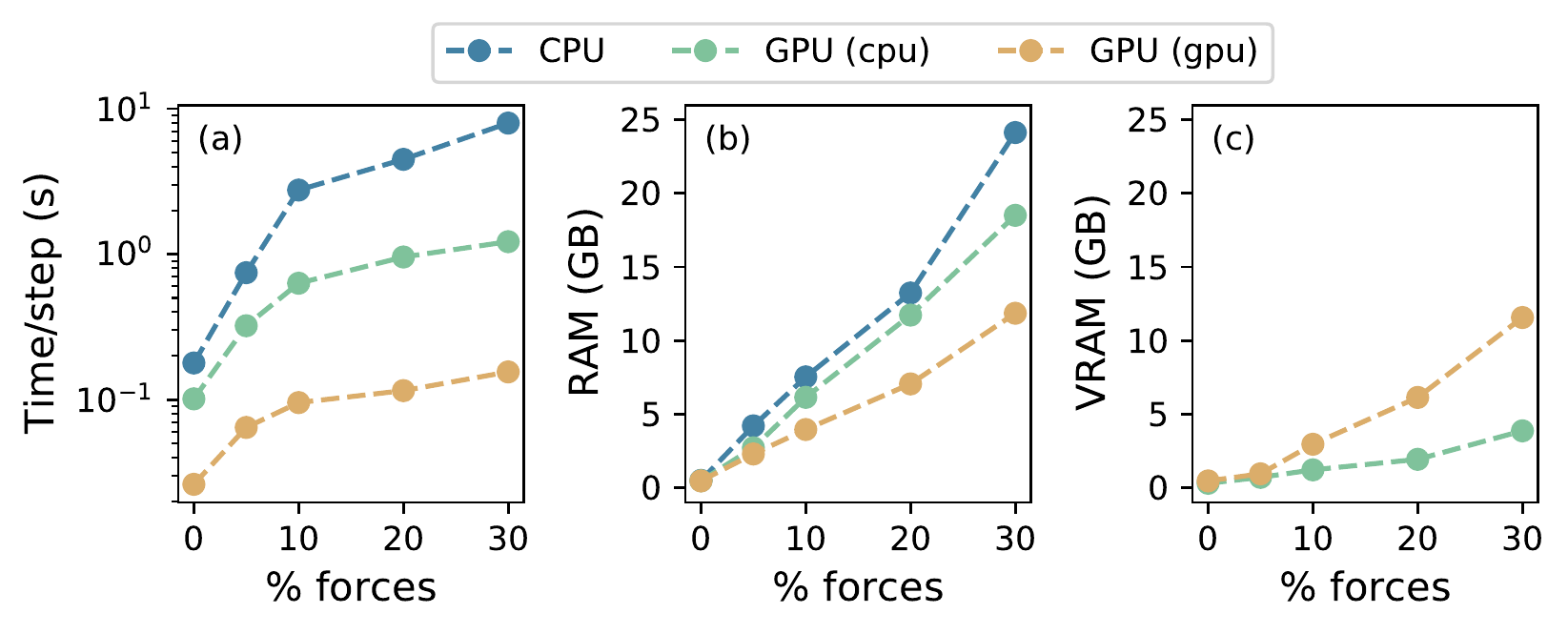}
    \caption{Resources used for training the neural network on both energies and forces of the \ch{TiO_2} database as a function of the fraction of force information included: (a) Time per training step, (b) CPU RAM memory and (c) GPU VRAM memory.}
    \label{fig3:F_scaling}
\end{figure}

Moreover, the number of forces included in the training examples heavily influences the computational requirements. This is shown in Fig. \ref{fig3:F_scaling}, where the training time, CPU RAM memory, and GPU VRAM memory for all three training strategies are displayed, this time as a function of the percentage of forces included. First, we note that including forces, even the smallest amount, increases at least 1 order of magnitude the time needed for training, which increases with the fraction of forces included. However, it is the memory that suffers the most with the addition of forces, rapidly increasing with it. This is somewhat mitigated by storing all the batch information in the CPU and only moving to the GPU the information one at a time. In any case, as the next section will show, in most applications including more than a small subset of the forces in the training set is not worth the computational cost.

\section{Benchmark on open data sets}

In this section, we will benchmark our code using several open databases, and compare them with the results obtained by their authors. Unless otherwise stated, the descriptor used has been the Chebyshev polynomials with the same parameters as in the reference article, and so is the chosen architecture. Here, we will use $\alpha=0.1$ to weigh the error for energy and force. Based on the analysis performed in the previous section, the prediction of energies is expected to be slightly less accurate than the one that would be achieved by fitting energies alone, but that will also result in more accurate force predictions.

\subsection{Titanium dioxide}

First, we summarize the results that have already been shown during this manuscript for the database of 7815 structures of several titanium dioxide phases developed by Artrith et al. and which was used to test the original ænet code in its first release \cite{aenet:artrith2016implementation}. Reference energy and forces were obtained from DFT calculations using the PBE exchange-correlation functional \cite{dft:perdew1996generalized}. In that work, the model was fitted only to the reference energies, and the descriptor used was the Behler-Parrinello symmetry functions\cite{mlp:behler2011atom}. The best fit resulted in an error of around 4 meV/atom in the energies, while the error on forces was not quantified in that work.

\begin{figure*}[ht!]
    \centering
    \includegraphics[width=0.8\textwidth]{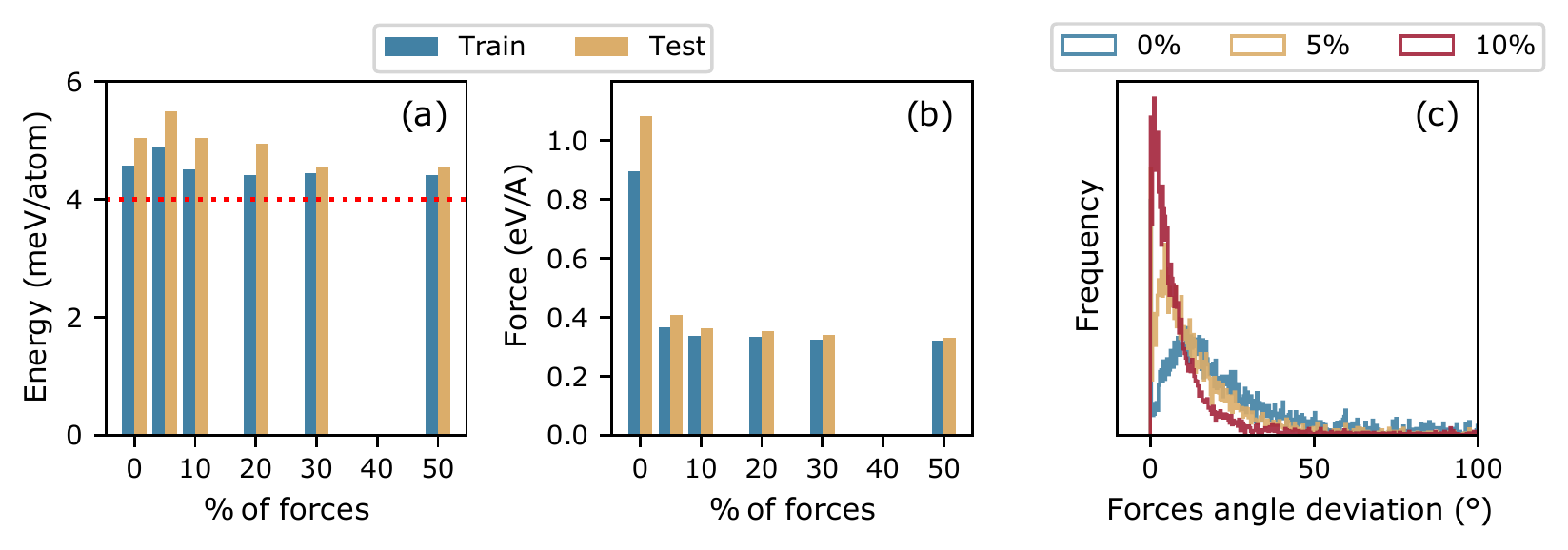}
    \caption{(a) Energy and (b) force RMSE of the \ch{TiO_2} data set for varying amount of force information. Blue for the training set and yellow for the validation set. The red dotted lines show the results obtained by the original authors of the data set. (c) Distribution of the error of the direction of the predicted forces, computed as the angle between the reference and the predicted forces.}
    \label{fig4:benchmark_TiO}
\end{figure*}

Our results are shown in Fig. \ref{fig4:benchmark_TiO}, where we include the errors in energies and forces for both the training and testing sets. As we have already mentioned before, the error in energies slightly increases with the number of forces, while the one for forces drastically diminishes. The red dotted line shows the error obtained in the original work, which is slightly better than the one we get here for the energy. This is due to the choice of the $\alpha$ parameter, a similar accuracy to that of the original article could be obtained by reducing the value of that parameter, but this would come at the cost of increasing the error of the force prediction (for example with $\alpha=0.01$ in Fig. \ref{fig3:F_scaling}).

On the other side, the error of the predicted absolute value of force converges to around 0.3 eV/\AA{} with only $20\%$ of forces included in the training. This error corresponds to around $2\%$ of the mean force of the structures on the data set ($\sim 20$ eV/\AA). Fig. \ref{fig4:benchmark_TiO}(c) shows the error in the direction of the predicted forces for different percentages. The addition of forces greatly improves its prediction, which is of great importance for MD simulations to be stable and accurate. Note that in cases where the absolute value of the force is smaller, the error in direction is more likely to be larger \cite{aenet:cooper2020efficient}.

\subsection{Liquid water}

Second, we consider a database composed of 9189 liquid water structures, each with 192 atoms, whose energies and forces were evaluated with the revised PBE functional and with the addition of the Grimme D3 van-der-Waals correction\cite{dft:grimme2010consistent}. This data set was used by Chen et al. to test the performance of the LAMMPS and TINKER interfaces of ænet \cite{aenet:chen2021aenet}. Water systems have been widely used as a benchmarking system for new developments in the field of MLPs \cite{water:bartok2013machine,water:cheng2019ab,water:quaranta2018structure,water:singraber2019library} due to their complexity, so many open databases can be found. This one here is more of a challenge than the one for \ch{TiO_2}, since the number of atoms in each structure is much higher, and therefore so is the number of reference forces to be fitted. In this case, the authors also fitted the model to the reference energies.

\begin{figure*}[ht!]
    \centering
    \includegraphics[width=0.8\textwidth]{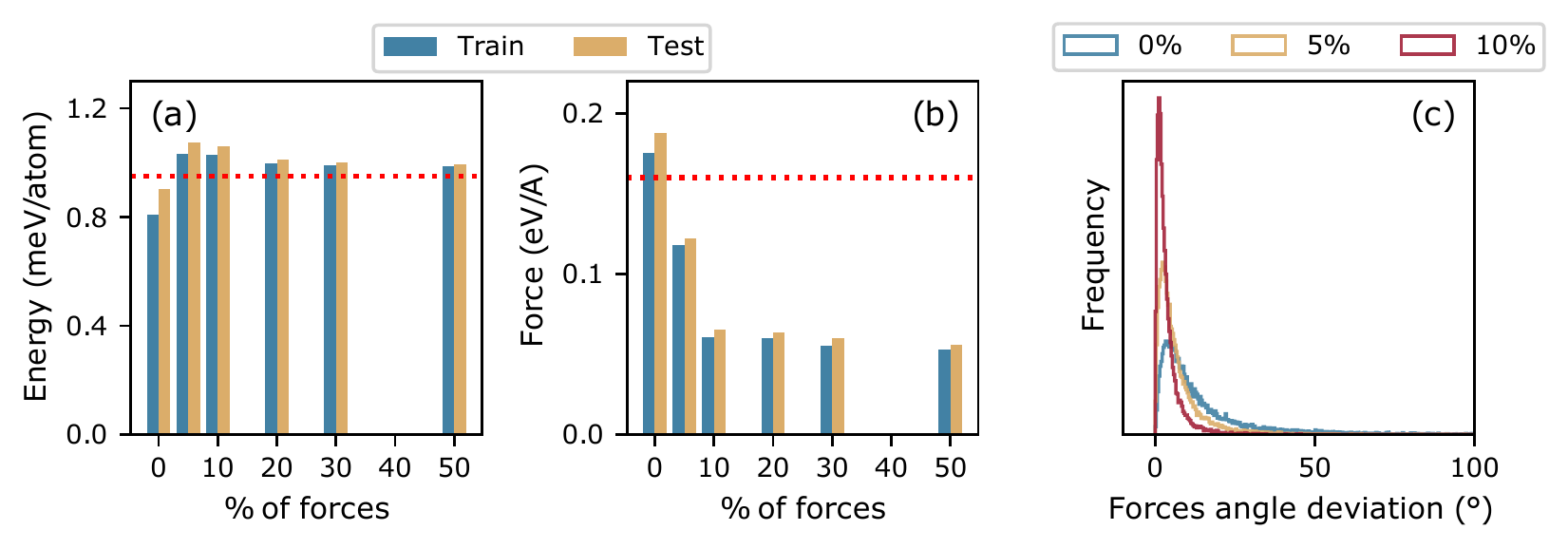}
    \caption{(a) Energy and (b) force RMSE of the \ch{H_2O} data set as a function of the percentage of forces included in the training stage. Blue for the training set and yellow for the validation set. The red dotted lines show the results obtained by the original authors of the data set. (c) Distribution of the error of the direction of the predicted forces.}
    \label{fig5:benchmark_H2O}
\end{figure*}

This time our results for energy training [Fig. \ref{fig5:benchmark_H2O}(a)] are in excellent agreement with the fitting of Chen et al., about 1 meV/atom. The error in forces is again lower in our fit, reducing  about $50\%$ with the smallest fraction of forces included. The absolute error of the forces cannot be directly compared to the results for the previous data set, but the relative error is around $1\%$ of the mean forces of the set, which is much lower in this case (around 2 eV/\AA) than that of \ch{TiO_2}. The most remarkable improvement comes from the angle deviation of forces, which becomes narrow around $0º$.

\subsection{Li-Mo-Ni-Ti oxide}

Our next example is a much more complex quaternary oxide, LMNTO, a database consisting of 2616 bulk structures, with 56 atoms each. Here the SCAN semilocal functional\cite{dft:sun2015strongly} was used to evaluate energies and forces. This reference data set was used by Cooper et al. to test their method for including force information via Taylor series expansions \cite{aenet:cooper2020efficient}, so we have the reference value of the errors for fitting to forces in addition to energies in this case. Moreover, being a system composed of 5 elements, it is a great example of the case in which the Chebyshev polynomials excel as atomic fingerprints, since the resulting descriptor size does not depend on the total number of elements\cite{mlp:artrith2017efficient}.

\begin{figure*}[ht!]
    \centering
    \includegraphics[width=0.8\textwidth]{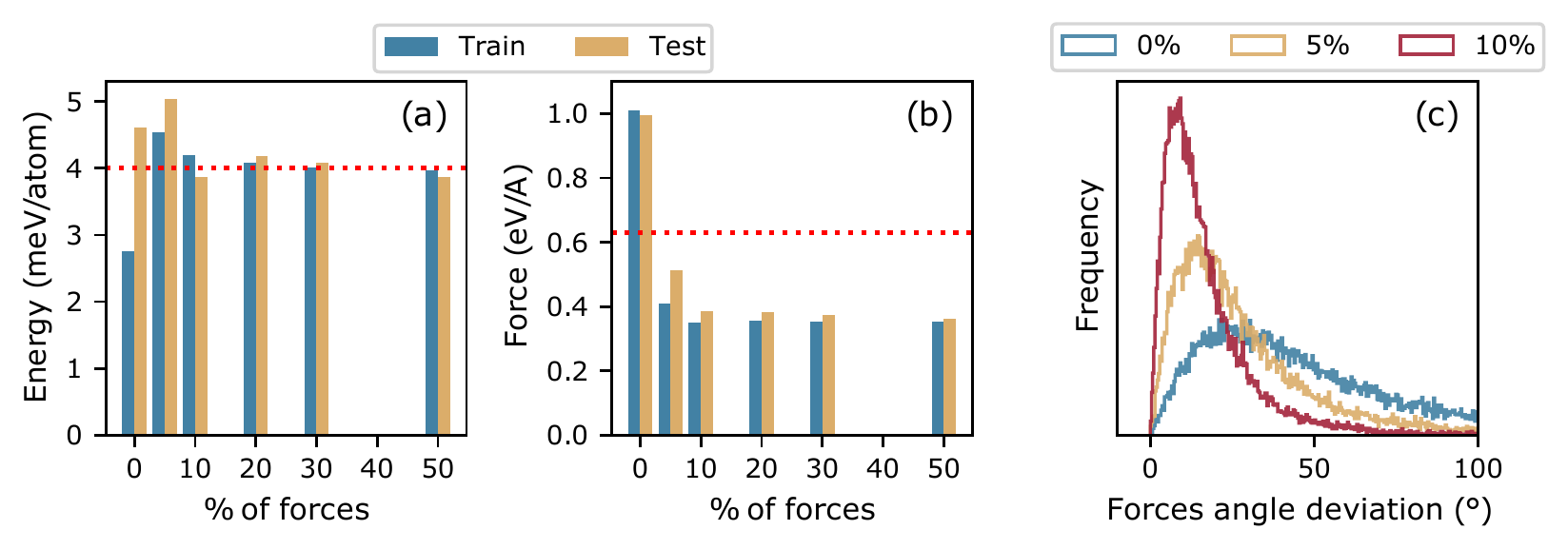}
    \caption{(a) Energy and (b) force RMSE of the \ch{LMNTO} data set as a function of the percentage of forces included in the training stage. Blue for the training set and yellow for the validation set. The red dotted lines show the results obtained by the original authors of the data set. (c) Distribution of the error of the direction of the predicted forces.}
    \label{fig6:benchmark_LMNTO}
\end{figure*}

Fig. \ref{fig6:benchmark_LMNTO} demonstrates that our models perform equally well in predicting energies compared to the reference models, but outperform them in terms of force prediction.  With no force information, our models have a higher error than that of Cooper et al., however, direct training on forces leads to improved performance over the Taylor series expansion. As in the two previous examples, the inclusion of forces initially increases energy error, but when sufficient force information is included, our models perform better than with only energy training. This highlights the benefit of incorporating forces in enhancing the generalization and transferability of the potentials.

\subsection{Amorphous \ch{Li_xSi} materials}

The last database that we will consider consists of about 45.000 structures of amorphous \ch{Li_xSi}, developed using a combination of density functional theory calculations, using the PBE functional, and evolutionary algorithms \cite{aenet:artrith2018constructing}. This set includes many phases with different stoichiometry, both bulk, surfaces and nanoparticles. In this case, only energies are included in the database, so force training is not possible, but it will still be useful to see the performance of our code in a large database with many atoms per structure.

The best fit using ænet-PyTorch yields an error of 6.5 meV/atom in the training set and 7.6 meV/atom in the testing set, which is as good as the reference fit (6.3 meV/atom in training and 7.7 eV/atom in testing). As for the resources, 10 GB of memory was needed to fit the energies of the whole set with 128 structures per batch.

\section{Conclusions}

The last decade has shown that MLPs will play an important role in the study of new and complex materials at the atomic scale, and this has created a huge demand for tools to efficiently train such potentials. Training on GPUs is the standard practice in most fields of machine learning, and here we presented an upgrade of the original ænet software to provide this capability. The ænet-PyTorch extension ensures that training the neural networks is no longer a bottleneck in the development of MLPs, even when accurate forces are required.

We demonstrated with different materials examples that ænet-PyTorch is efficient, particularly when training only on energies. The CPU version of the code is as fast as the original ænet implementation, and the GPU implementation reduces training time by one to two orders of magnitude. Due to its compatibility with the ænet package, we expect this extension to have great synergy with the other features available in ænet, such as including force information in the training via a Taylor series expansion\cite{aenet:cooper2020efficient}.

If, on the other hand, direct training using atomic forces is desired, this is now feasible with the ænet-PyTorch code. We demonstrated that directly including force information in the training process is possible with ænet-PyTorch thanks to the computationally efficient GPU implementation. Nonetheless, we strongly recommend users to include only a small fraction of forces in the training, since our benchmarks demonstrated that accurate models can be obtained by including between $5\%$ to $20\%$ of the force information.

\begin{acknowledgments}
This work has been supported by the ‘‘Departamento de Educación, Política Lingüística y Cultura del Gobierno Vasco’’ (IT1458-22), the ‘‘Ministerio de Ciencia e Innovación" (PID2019-106644GB-I00), and the Project HPC-EUROPA3 (INFRAIA-2016-1-730897), with the support of the EC Research Innovation Action under the H2020 Programme. The authors thank for technical and human support provided by SGIker (UPV/EHU/ ERDF, EU), and the Duch National e-Infrastructure and the SURF Cooperative for computational resources (National Supercomputer Snellius). J.L.-Z. acknowledges financial support from the Basque Country Government (PRE\_2019\_1\_0025). N. A. gratefully acknowledges funding from Bayer Life Science through the !AIQU project.
\end{acknowledgments}

\section*{Data availability}

ænet-PyTorch is open source and free for everyone to use and customize, as is the ænet package. The ænet-PyTorch code can be obtained from \url{https://github.com/atomisticnet/aenet-PyTorch}. Being written purely in Python and PyTorch, we believe that this code can be easily used for prototyping new techniques based on PyTorch features, such as custom loss functions, learning rate schedulers, and dropout layers to reduce overfitting.

\section*{References}
\bibliography{pytorch-aenet}

\end{document}